\documentclass[twocolumn,aps,pra,showpacs,amsmath,amssymb,superscriptaddress, longbibliography]{revtex4-1}

\usepackage{physics}
\usepackage{braket}
\usepackage{graphicx} 
\usepackage{float}
\usepackage[caption=false, labelformat=simple]{subfig}
\usepackage{dcolumn}
\usepackage{csquotes}
\usepackage[colorlinks=true,linkcolor = {blue}, citecolor = {blue}, urlcolor = {blue}]{hyperref}
\usepackage{xcolor}

\begin{document}

\title{Coupled-channels method for the scattering hypervolume in ultracold atomic three-body collisions}

\author{P.J.P. Kersbergen}
\email[Corresponding author:  ]{p.j.p.kersbergen@tue.nl}
\affiliation{Department of Applied Physics and Science Education, Eindhoven University of Technology, P. O. Box 513, 5600 MB Eindhoven, The Netherlands}
\author{J. van de Kraats}
\affiliation{Department of Applied Physics and Science Education, Eindhoven University of Technology, P. O. Box 513, 5600 MB Eindhoven, The Netherlands}
\author{D.J.M. Ahmed-Braun}
\affiliation{Theory of Quantum and Complex Systems, Physics Department, Universiteit Antwerpen, B-2000 Antwerpen, Belgium}
\author{S.J.J.M.F. Kokkelmans}
\affiliation{Department of Applied Physics and Science Education, Eindhoven University of Technology, P. O. Box 513, 5600 MB Eindhoven, The Netherlands}
\date{\today}

\begin{abstract}
We introduce a novel coupled‑channels method for elastic three‑body scattering in systems of identical bosonic alkali‑metal atoms. The approach relies on the numerically exact two‑body off‑the‑energy‑shell transition matrix, constructed from realistic multichannel molecular interaction potentials that support many bound states. By rigorously accounting for this off‑shell structure, the method captures both the short‑range physics as well as multichannel couplings characteristic of alkali‑metal potentials without resorting to model pseudopotentials. The central output is the complex three‑body scattering hypervolume---the three‑body analogue of the two‑body scattering length---which we obtain with controlled and verifiable numerical accuracy. As a realistic benchmark, we apply our framework to spin‑polarized potassium-39, performing full coupled‑channels three‑body scattering calculations and extracting the hypervolume over experimentally relevant conditions. The method is general and transferable to other atomic species and interaction models featuring deep molecular potentials with an arbitrarily large number of bound states.
\end{abstract} 

\maketitle

\section{Introduction}\label{Intro}
Ultracold atomic gases offer an exceptionally tunable platform for exploring interacting quantum matter. A key tool is the use of magnetic Feshbach resonances to control the s-wave scattering length $a$, the low‑energy measure of effective two‑body interactions in elastic atomic collisions, over a wide range of values, including changes in sign and magnitude \cite{Feshbach1, Feshbach2, chinFeshbach, taylor}. This tunability enables experimental access to weak‑ and strong‑coupling regimes and underpins the use of ultracold gases as quantum simulators for phenomena spanning condensed‑matter to nuclear physics \cite{condensedsim}. For example the Bardeen–Cooper–Schrieffer (BCS) pairing, relevant to conventional superconductors, and superfluid neutron matter in neutron‑star crusts \cite{supercond, neutronstars}.\\

Equilibrium and dynamical properties of dilute quantum gases are governed by few‑body scattering processes. This allows for a bottom‑up approach of the many‑body physics: one characterizes the microscopic interactions and then builds up the emergent properties of the many‑particle system \cite{ColussiThesis, ManybodyBloch}. At lowest order, the gas is described by two‑body physics, parameterized by the s-wave scattering length. Already at this level, mean field theory captures a rich set of quantum phases and crossovers, including Bose–Einstein condensation (BEC) in bosonic gases and the BEC–BCS crossover in strongly interacting Fermi gases \cite{BCS-BEC, ManybodyBloch}.

Effective three‑body interactions represent the next level of complexity beyond two‑body physics. In the low‑energy expansion, three‑body effects are parametrized by the three‑body scattering hypervolume $D$. The real part $\text{Re}(D)$ quantifies the effective elastic three‑body interaction strength, directly analogous to $a$ for two‑body scattering, and can modify the equation of state and stability of quantum gases. In particular, theoretical studies have demonstrated that repulsive three-body interactions can stabilize single-component bosonic systems against collapse and enable the formation of homogeneous quantum droplets \cite{bulgacdrops, bedaquedrops, zwergerdrops, PetrovControl}, distinct from the self-bound droplets found in binary mixtures \cite{dropletmixture}. Elastic three‑body terms are also predicted to influence the collapse dynamics of attractive BECs \cite{cambridgepaper, Collapsedynamics}, although their quantitative impact is typically subdominant and experimentally elusive. When interactions become strong, three‑body scattering gives rise to non-perturbative few‑body phenomena, most prominently Efimov physics \cite{efimov1, efimov2, EfimovReview, efimovevidence}. The imaginary part $\text{Im}(D)$ describes inelastic three‑body processes and is directly connected to the total three‑body recombination rate \cite{BraatenReport, Zhu+Tan, BookSchmid}, where three atoms collide to form a molecule and a free atom, releasing enough kinetic energy for both to escape the confinement. This process is a primary limitation on the lifetime of ultracold gases and BECs \cite{GreeneFewBody, RecombinationMeasurement, threebodyInflation, threebodyJasper, MultichannelEffectsEfimov}.

The bottom-up approach is evident in the low-density expansion of the ground state energy density $\mathcal{E}$ of a weakly-interacting BEC with density $n$, given as \cite{PaperTan, scathypPaul}
\begin{equation}\label{eq:BECenergy}
\begin{split}
\mathcal{E}=&\frac{2\pi\hbar^2n^2a}{m}\left(1+\frac{128}{15\sqrt{\pi}}\sqrt{na^3}+\left[\frac{8(4\pi-3\sqrt{3})}{3}\ln{(na^3)}\right.\right.\\
&\left.\left.+\frac{\pi r_s}{a}+\frac{D}{12\pi a^4}+118.5\right]na^3+...\right),
\end{split}
\end{equation}
where the dots correspond to higher-order corrections in terms of the diluteness parameter $na^3$ and $r_s$ is the effective range parameter. The $\sqrt{na^3}$ correction is the Lee-Huang-Yang (LHY) two-body correction \cite{LHYcorrection}. At the next order, the logarithmic $\ln{(na^3)}$ correction appears, calculated by Wu \cite{Wucorrection}. This correction arises from the coupling between short-range interaction physics and long-range collective modes, marking the onset of many-body correlations \cite{Wucorrection, Pinescorr}. The first contribution that is entirely decoupled from two-body scattering, and thus independent of $a$, quantifies genuine short-range three-body interactions \cite{PaperTan}, and is characterized by the three-body scattering hypervolume $D$.\\

From a theoretical perspective, computing quantitatively reliable three‑body parameters for alkali‑metal atoms is challenging. The underlying two‑body interactions are multichannel in nature and governed by deep molecular potentials with many bound states. Therefore, an accurate treatment requires the off‑shell two‑body transition matrix derived from realistic potentials, rather than model pseudopotentials or single‑channel approximations. This off‑shell structure controls short‑range physics and channel couplings that feed directly into three‑body scattering and hence into both $\text{Re}(D)$ and $\text{Im}(D)$. These considerations motivate methods that (i) incorporate realistic multichannel two‑body physics at the off‑shell level and (ii) deliver the three‑body scattering hypervolume $D$ with controlled numerical accuracy.

To date, most determinations of the hypervolume have been restricted to simplified model interactions, including hard‑sphere \cite{PaperTan}, Gaussian  \cite{Zhu+Tan}, square‑well \cite{scathypPaul}, and shallow van der Waals potentials \cite{vdWuniversality}. Two main strategies have been pursued. Firstly, position‑space approaches match the hyperspherical three‑body wavefunction to known asymptotic solutions \cite{PaperTan, Zhu+Tan, IncaoFewBody, IncaoRecom}. Unfortunately, these approaches do not accommodate the inclusion of internal spin degrees of freedom, despite their known, sometimes dramatic, impact on three‑body observables \cite{LiSpinExchange, threebodyJasper, threebodyInflation, ThomasMultiChannel}. Momentum‑space approaches, by contrast, solve the Alt–Grassberger–Sandhas (AGS) equations using a Weinberg expansion of the two‑body transition matrix (referred to here as the \textit{plane‑wave method}) \cite{PaulFiniteRange, PaperWeinberg}, as implemented in Refs.~\cite{scathypPaul, vdWuniversality}. However, momentum‑space treatments face delicate subtractions to remove divergences in the real part of the three‑body scattering amplitude \cite{scathypPaul, IncaoFewBody, energyexpansion, NietoCorrection}, which makes numerical precision paramount. While the plane‑wave method can yield accurate values of $D$ for shallow van der Waals interactions, its convergence with increasing potential depth is too slow for applications to the deep molecular potentials characteristic of alkali‑metal atoms \cite{PaulFiniteRange}.\\

In this work, we introduce a new coupled‑channels method for elastic three‑body collisions of identical bosonic alkali‑metal atoms that directly uses the off‑shell two‑body transition matrix obtained from realistic multichannel molecular potentials. The framework rigorously treats internal spin structure in a full coupled basis and overcomes the limitations of the plane‑wave approach for deep potentials. It yields the complex-valued hypervolume $D$ with verifiable numerical accuracy. Methodologically, our approach builds on a recently developed DVR‑based scheme that has enabled precise three‑body recombination calculations in multichannel alkali systems (referred to as the \textit{DVR-method}) \cite{threebodyInflation, threebodyJasper, MultichannelEffectsEfimov, LiSpinExchange, ThomasMappedGrid, ThomasMultiChannel}. Here we extend this toolbox to make it suitable for elastic scattering and the extraction of $D$ in realistic alkali‑metal settings. In doing so, we substantially widen the scope of quantitative three‑body modeling in ultracold atomic gases.\\

This paper is organized as follows. In Sec.~\ref{Method-theory} we introduce the Hamiltonian for three interacting alkali‑metal atoms and the three‑body basis used to formulate the scattering problem. In Sec.~\ref{Method-scathyp} we cast the problem in terms of the AGS equations for the three‑body transition matrix and relate these to the scattering hypervolume. Sec.~\ref{Method-2body} describes the highly accurate determination of the off‑shell two‑body transition matrix at low energies. Sec.~\ref{Method-3body} presents the numerical implementation and analyzes convergence with respect to relevant parameters. In Sec.~\ref{Method-benchmarking}, we benchmark the cancellation of divergent contributions by quantitatively comparing results for a shallow single‑channel van der Waals potential with the plane‑wave method of Ref.~\cite{vdWuniversality}. Sec.~\ref{Results} demonstrates the capabilities of our method by presenting results for spin‑polarized $^{39}\mathrm{K}$ and analyzing how multichannel effects modify universal predictions. We conclude with an outlook for future research.

\section{Method}\label{Method}
\subsection{Three-body Hamiltonian and basis}\label{Method-theory}
The three-body system is described by Jacobi coordinates, where the system is partitioned into a pair (forming a dimer subsystem) and a third particle \cite{Glöckle}. This gives two relative momenta; $\mathbf{k}_\alpha$ and $\mathbf{p}_\alpha$, where $\alpha=1,2,3$ is the index that corresponds to the third particle. This labeling will be used throughout this paper. The momentum $\mathbf{k}_\alpha$ is the relative momentum of the pair, whilst $\mathbf{p}_\alpha$ is the relative momentum of particle $\alpha$ with respect to the center of mass of the pair \cite{Glöckle}. The atoms are assumed to interact via pairwise interaction potentials $V_\alpha$, which act purely on the pair, thus leaving particle $\alpha$ to spectate. 

Long before and long after a scattering process, the atoms may exist in different so-called configuration channels, determining the asymptotic structure of the three-body state. In the break-up channel, all particles asymptote to non-interacting single-particle states, i.e. all particles are asymptotically free. Conversely, in a fragmentation channel, two particles remain asymptotically bound in a dimer state, while the third particle is free. We label the break-up channel with index 0, and label the three fragmentation channels by the third particle index $\alpha$.

In this work we employ the same three-body partial-wave basis as Ref.~\cite{threebodyJasper}; $\ket{k_\alpha,l_\alpha,m_{l_\alpha}}\otimes\ket{p_\alpha,\lambda_\alpha,m_{\lambda_\alpha}}$, where $l_\alpha$ is the orbital angular momentum quantum number of the dimer subsystem and $\lambda_\alpha$ is the atom-dimer angular momentum quantum number. Since the three-body system is completely rotationally symmetric, it is useful to couple the dimer and atom-dimer angular momenta to produce a total angular momentum $\mathbf{L}=\mathbf{l}_\alpha+\boldsymbol{\lambda}_\alpha$, with projection $M_L$. A given basis state of the three-body system is then written as $\ket{k_\alpha p_\alpha(l_\alpha\lambda_\alpha)LM_L}$. Going forward we will restrict the three-body basis to those states with $L=M_L=0$, as these states provide the dominant contribution to the three-body physics at low energies \cite{threebodythreshold}. This restriction enforces that $l_\alpha=\lambda_\alpha$, such that we retain only the states $\ket{k_\alpha p_\alpha(l_\alpha l_\alpha)00}$. \\

When modeling the spin-structure of alkali-metal atoms, we note that these atoms only have one valence electron and hence have a total electron spin of $s=1/2$. The nuclear spin $i$ is dependent on the type of nucleus (for $^{39}\mathrm{K}$ the nuclear spin is $i=3/2$) \cite{Pethick}. The spin-spin coupling of the nucleus and valence electron is represented by the hyperfine interaction,
\begin{equation}\label{eq:hfinteraction}
H_{hf}=A_{hf}\mathbf{s\cdot i},
\end{equation}
where the hyperfine constant $A_{hf}$ depends on the atomic species and has typical values of hundreds to thousands of MHz \cite{Pethick}. In the presence of an external magnetic field $\mathbf{B}$, the Zeeman interaction further provides a contribution $H_Z$ to the spin Hamiltonian ($H_{\mathrm{spin}} =H_{hf}+H_{Z}$), where
\begin{equation}\label{eq:Zeeman}
H_Z=\gamma_n\mathbf{i\cdot B}+\gamma_e\mathbf{s\cdot B},
\end{equation}
with $\gamma_n$ and $\gamma_e$ the nuclear and electronic gyromagnetic ratios \cite{Pethick}. Under application of a magnetic field, the degeneracy of hyperfine states $\ket{f,m_f}$ is lifted (with total spin $\mathbf{f}=\mathbf{s}+\mathbf{i}$), such that the quantum number $m_f$ now labels a unique state within each hyperfine manifold. Conventionally, each single-particle eigenstate is labeled by the hyperfine state they adiabatically connect to for $|\mathbf{B}| \to 0$ \cite{LiSpinExchange}.

When the interatomic distance between two atoms becomes small, the valence electrons overlap, projecting the two-body state into singlet (total spin $\mathbf{S}=\mathbf{s}_1+\mathbf{s}_2=0$) and triplet ($\mathbf{S}=1$) molecular projections. At these short distances, the interaction potential $V_\alpha$ decomposes correspondingly in singlet and triplet components;
\begin{equation}\label{eq:potentialproj}
V_\alpha=V_\alpha^0\mathcal{P}_\alpha^0+V_\alpha^1\mathcal{P}^1_\alpha,
\end{equation}
where $\mathcal{P}^S_\alpha$ projects the pair into the spin state with spin $S$. Since this projection does not commute with the hyperfine Hamiltonian, the potential is not diagonal in the basis of hyperfine states. This induces spin-exchange couplings and necessitates a multichannel description of the two-body interaction \cite{MiesMultChan1, MiesMultChan2}. To model a specific system of alkali-metal atoms in detail, the singlet and triplet molecular interaction potentials $V_\alpha^{0,1}$ are computed from available empirical data, such as loss measurements of Feshbach resonance positions and spectroscopy of molecular binding energies \cite{PotassiumResonance1, d-waveResonancePotassium, MixtureResonancePotassium, TiemannPot, JuliennePot}. These potentials consist of a non-universal short-range repulsion and a universal long-range attraction that obeys
\begin{equation}\label{eq:longrangepot}
V(r)\sim-\frac{C_6}{r^6}-\frac{C_8}{r^8}-\frac{C_{10}}{r^{10}},
\end{equation}
where $C_6$, $C_8$ and $C_{10}$ are the strengths of each type of interaction and are determined from fitting experimental observations, as done in Ref.~\cite{TiemannPot}. The contribution that decays the slowest is the well known van der Waals interaction, where the parameter $C_6$ determines the van der Waals range $r_{\mathrm{vdW}}=(mC_6/\hbar^2)^{1/4}/2$, which sets the characteristic length scale of the problem. These potentials are deep (on the order of THz) as compared to hyperfine splitting and contain many bound states.

Because of spin-exchange in the interactions, the only rigorously conserved quantity in the three-body system is the total magnetic spin projection $M_f=\sum_{j=1}^3m_{f_j}$, which determines the accessible hyperfine states and therefore the numerical complexity of the coupled system.  

The eigenstates of the complete spin-Hamiltonian, given as the sum of the single-particle Hamiltonians in Eqs.~\eqref{eq:hfinteraction} and \eqref{eq:Zeeman}; $\sum_{j=1}^3H_{hf_j}+H_{z_j}$, will be referred to as channel states, given as $\ket{C_\alpha,c_\alpha}$, with $c_\alpha$ the spin channel of the third free particle and $C_\alpha$ the spin channel of the dimer subsystem, with energies $E_{C_\alpha,c_\alpha}=\varepsilon_{C_\alpha}+\varepsilon_{c_\alpha}$. Additionally, the total dimension of the dimer basis can be reduced by demanding that the allowed states are symmetric under particle exchange, where we also correct for the spatial parity of odd and even angular momentum quantum numbers $l$, resulting in
\begin{equation}\label{eq:symmetry}
\ket{\bar{C}_\alpha}=\frac{1+(-1)^lP_\alpha}{\sqrt{2(1+\delta_\alpha)}}\ket{C_\alpha},
\end{equation}
where $P_\alpha$ denotes the permutation operator of exchanging the two paired particles and $\delta_\alpha$ is unity when the two single-particle states are identical. Using the channel states, we define the complete three-body basis as $\ket{k_\alpha p_\alpha(l_\alpha l_\alpha)00}\otimes\ket{\bar{C}_\alpha,c_\alpha}$. For our reference energy, we set $E=0$ at the three-body scattering threshold, given by the lowest channel energy. For notational brevity, we will suppress the index $\alpha$ on momenta and quantum numbers in the remainder of this paper, because the appropriate index can always be traced back from the surrounding operators and equations.  

\subsection{Scattering hypervolume}\label{Method-scathyp}
The scattering hypervolume is extracted from the low-energy limit of the elastic three-body transition operator in the break-up channel, $U_{00}(z)$ \cite{PaperTan}. This transition operator can be calculated from the Alt-Grassberger-Sandhas (AGS) equations, given as \cite{AGSpaper}
\begin{equation}\label{eq:elasticAGS}
U_{00}=\sum_{\alpha=1}^3\mathcal{T}_\alpha G_0U_{\alpha0},
\end{equation}
where for brevity the energy arguments $(z)$ are suppressed. Here, $G_0$ is the free three-body Greens function and $\mathcal{T}_\alpha$ is the two-body transition matrix of a pair in channel $\alpha$. $U_{\alpha 0}$ is the transition element of inelastic scattering from a fragmentation to break-up channel. 

Due to the symmetry property of identical bosons, the following transformation will prove to be very convenient: $\breve{U}_{\alpha 0}=\mathcal{T}_\alpha G_0U_{\alpha 0}(1+P)$, with $P=P^++P^-$ the permutation operators that cycles particle indices clockwise ($P^+$) or anticlockwise ($P^-$). This transformation allows us to reformulate Eq.~\eqref{eq:elasticAGS} into
\begin{equation}\label{eq:elasticAGStrans}
U_{00}=\frac{1}{3}U_{00}(1+P) = \frac{1}{3}\sum_{\alpha=1}^3\breve{U}_{\alpha0}.
\end{equation}
The inelastic transition operators $\breve{U}_{\alpha0}$ are also calculated from an AGS equation, 
\begin{equation}\label{eq:inelasticAGStrans}
\breve{U}_{\alpha0}=\mathcal{T}_\alpha(1+P)+\mathcal{T}_\alpha G_0P\breve{U}_{\alpha 0}.
\end{equation}
In similar fashion to the two-body scattering length, the three-body scattering hypervolume can be computed from the zero momentum and zero energy limit of the on-shell element of the three-body transition matrix $\braket{kp(ll)00,\bar{C}c|\breve{U}_{\alpha0}(z)|00(00)00,\bar{C}_{\mathrm{in}}c_{\mathrm{in}}}$, corresponding to scattering from an initial zero-momentum state to the continuum. This element at $k=0$ has the following low atom-dimer momentum expansion \cite{scathypPaul, PaperTan} \begin{equation}\label{eq:divergingtransampl}
\begin{split}
&\braket{0p(ll)00;\bar{C}c|\breve{U}_{\alpha 0}(0)|00(00)00;\bar{C}_{\mathrm{in}}c_{\mathrm{in}}} =\\
&\left[\frac{a}{2\pi^2m\hbar}\delta(p)-\frac{3a^2}{2\pi^4m\hbar^4}\frac{1}{p^2}\right.+\frac{\left(4\pi-3\sqrt{3}\right)a^3}{4\pi^4m\hbar^4}\frac{1}{p}\\
&+\left.\frac{\left(4\pi-3\sqrt{3}\right)a^4}{\pi^5m\hbar^4}\ln{\left(p|a|\right)}\right]\delta_{l0}\delta_{cc_{\mathrm{in}}} \\&+\braket{0p(ll)00;\bar{C}c|\breve{U}_{\alpha 0}^{\mathrm{nd}}(0)|00(00)00;\bar{C}_{\mathrm{in}}c_{\mathrm{in}}}\\
&=\braket{0p(ll)00;\bar{C}c|\breve{U}_{\alpha 0}^{\mathrm{div}}(0)+\breve{U}_{\alpha 0}^{\mathrm{nd}}(0)|00(00)00;\bar{C}_{\mathrm{in}}c_{\mathrm{in}}}.
\end{split}
\end{equation}
Eq.~\eqref{eq:divergingtransampl} contains terms that diverge at low atom-dimer momentum and terms that are non-diverging (they vanish or go to a constant), these two types are collected in matrix elements $\breve{U}_{\alpha 0}^{\mathrm{div}}$ and $\breve{U}_{\alpha 0}^{\mathrm{nd}}$ respectively. Note that the factor $\delta_{l0}\delta_{cc_{\mathrm{in}}}$ cause the diverging contributions to only appear in the incoming s-wave channel.

Crucially, the scattering hypervolume, appearing as a correction to the BEC ground state in the dilute and weakly-interacting limit (see Eq.~\eqref{eq:BECenergy}), is precisely the non-divergent part of the three-body transition operator \cite{PaperTan, Zhu+Tan}. So, using Eq.~\eqref{eq:elasticAGStrans} to convert $\breve{U}_{\alpha 0}$ to $U_{00}$ and following the formal definition of Ref.~\cite{PaperTan} we can find the scattering hypervolume $D$ by taking the limit $k,p,z\to0$, such that explicitly
\begin{widetext}
\begin{equation}\label{eq:scathypvol}
\frac{1}{(2\pi)^6}\frac{D}{m\hbar^4}=\frac{1}{3}\sum_{\alpha=1}^3\lim_{k,p\to0}\left(\braket{kp(ll)00;\bar{C}c|\breve{U}_{\alpha 0}^{\mathrm{nd}}(0)|00(00)00;\bar{C}c}\left.+\frac{k^2+\frac{3}{4}p^2}{p^2}\frac{3a}{2\pi^2}\frac{\partial^2}{\partial k^2}\braket{k;\bar{C}|\mathcal{T}_\alpha(0)|0;\bar{C}}\right|_{k=0}\right).
\end{equation}
\end{widetext}
In order to compute the three-body transition matrix, the two-body transition matrix $\mathcal{T}_\alpha$ is required. It is diagonalized to obtain the spectral expansion \cite{PaulFiniteRange, sitenko}
\begin{equation}\label{eq:tmatrixexpansion}
\braket{k'l;\bar{C}|\mathcal{T}_\alpha(z)|kl;\bar{C}}=\sum_n\braket{k';\bar{C}|\chi_{l}^n(z)}\tau_{l}^n(z)\braket{\chi_l^n(z)|k;\bar{C}},
\end{equation}
with $\braket{k;\bar{C}|\chi_{l}^n(z)}$ the $n$-th form factor and $\tau_{l}^n(z)$ the $n$-th eigenvalue. Here $n$ is the amount of separable terms that we take into account before truncating the spectral expansion, turning it into a separable approximation to the formal transition matrix \cite{sitenko}. The calculation of the form factors and eigenvalues of the separable approximation are delegated to the next section. 

Then, to compute the non-divergent part of the transition matrix, we transform Eq.~\eqref{eq:inelasticAGStrans} by inserting $\breve{U}_{\alpha 0}=\breve{U}_{\alpha 0}^{\mathrm{div}}+\breve{U}_{\alpha 0}^{\mathrm{nd}}$ to isolate the diverging contribution such that we explicitly find, 
\begin{equation}\label{eq:nondivtransampl}
\breve{U}_{\alpha0}^{\mathrm{nd}}=-\breve{U}_{\alpha 0}^{\mathrm{div}}+\mathcal{T}_\alpha(1+P)+\mathcal{T}_\alpha G_0P(\breve{U}_{\alpha 0}^{\mathrm{div}}+\breve{U}_{\alpha 0}^{\mathrm{nd}}).
\end{equation}
By inserting complete sets of the three-body basis states (as introduced in Sec.~\ref{Method-theory}) between all the operators and using the separable expansion of the two-body T-matrix, we can transform this operator equation into a one-dimensional integral equation over the atom-dimer momentum $q$ \cite{threebodyJasper}
\begin{equation}\label{eq:intequation}
\begin{split}
&\mathcal{U}^n_{lc}(p)=\mathcal{I}^n_{lc}(p)\\
&+ 8\pi\hbar^3\sum_{l',c',n'}\int_0^\infty q^2\mathcal{Z}^{nn'}_{ll';cc'}(p,q)\tau_{l'}^{n'}(Z_{c'}(q))\mathcal{U}^{n'}_{l'c'}(q)dq,
\end{split}
\end{equation}
where $\mathcal{U}^n_{lc}(p)$ is defined as 
\begin{equation}\label{eq:operatortointegral}
\begin{split}
&\braket{kp(ll)00;\bar{C}c|\breve{U}^{\mathrm{nd}}_{\alpha 0}(0)-\mathcal{T}_\alpha(1+P)|00(00)00;\bar{C}c}\\
&=\sum_n\braket{k;\bar{C}|\chi_l^n(Z_c(p))}\tau_l^n(Z_c(p))\mathcal{U}_{lc}^n(p),
\end{split}
\end{equation}
such that we separate the dependence on the dimer momentum $k$ and dimer spin-channel $\bar{C}$. The energy $Z_c(p)=z-\frac{3}{4}\frac{\hbar^2p^2}{m}-\varepsilon_c$ corresponds to the non-interacting energy of the dimer.

The inhomogeneous term $\mathcal{I}^n_{lc}(p)$ in Eq.~\eqref{eq:intequation} contains the diverging terms $\breve{U}_{\alpha 0}^{\mathrm{div}}$ that allow us to directly compute the matrix elements of $\breve{U}_{\alpha 0}^{\mathrm{nd}}$,
\begin{equation}\label{eq:inhomterm}
\begin{split}
&\mathcal{I}^n_{lc}(p)=-\mathcal{U}^{\mathrm{(div)}}(p)\\
&+6\sum_{c',n'}\mathcal{Z}_{l0;cc'}^{nn'}(p,0)\tau_0^{n'}(Z_{c'}(0))\braket{\chi_0^{n'}(Z_{c'}(0))|0;\bar{C}}\\
&+ 8\pi\hbar^3\int_0^\infty q^2\mathcal{Z}^{n0}_{l0;cc}(p,q)\tau_{0}(Z_{c}(q))\mathcal{U}^{\mathrm{(div)}}(q)dq,
\end{split}
\end{equation} 
with $$\mathcal{U}^{\mathrm{(div)}}(p)=\frac{\braket{0p(00)00;\bar{C}c|\breve{U}^{\mathrm{div}}_{\alpha 0}(0)|00(00)00;\bar{C}c}}{\braket{0;\bar{C}|\chi_0(Z_c(p))}\tau_0(Z_c(p)))}.$$ We have limited the separable expansion in the divergent channel to just a single term, labeled with $n = 0$. As we will show in the next section, this will be sufficient for our expansion of the two-body transition matrix in the low-momentum regime. 

The kernel function in Eq.~\eqref{eq:intequation} is $\mathcal{Z}^{nn'}_{ll';cc'}(p,q)$ and is defined as
\begin{equation}\label{eq:Kernelfunction}
\begin{split}
&\mathcal{Z}^{nn'}_{ll';cc'}(p,q)=\frac{1}{2}\sum_{\bar{C}'}\sum_{\bar{C}''}\int_{-1}^1\mathfrak{B}_{ll'}(p,q,x)\\
&\times\frac{\braket{\chi_l^n(Z_c(p))|\pi(q,p,x);\bar{C}'}\braket{\pi(p,q,x);\bar{C}''|\chi_{l'}^{n'}(Z_{c'}(q))}}{z-\frac{\hbar^2}{m}(q^2+p^2+qpx)-E_{\bar{C}'c}}\\
&\times\braket{l;\bar{C}',c|P_{+}^s|l';\bar{C}'',c'}dx,
\end{split}
\end{equation}
where $E_{\bar{C}'c}$ is the energy of the relevant three-body channel and where the matrix element $\braket{l;\bar{C}',c|P_{+}^s|l';\bar{C}'',c'}$ corresponds to the clockwise cycling of spin-indices \cite{threebodyJasper, MultichannelEffectsEfimov}. We use the definition $\pi(p,q,x)=\sqrt{p^2+\frac{1}{4}q^2+pqx}$ and the function $\mathfrak{B}_{ll'}(p,q,x)$ to express the coupling of angular momenta via
\begin{equation}\label{eq:recouplingfunction}
\begin{split}
\mathfrak{B}_{ll'}&(p,p',x)=(-1)^{l'}\sqrt{(2l+1)(2l'+1)}\\
&\times P_l\left(\frac{\frac{1}{2}p^2+pp'x}{p\pi(p',p,x)}\right)P_{l'}\left(\frac{\frac{1}{2}p'^2+pp'x}{p'\pi(p,p',x)}\right),
\\[1mm]
\end{split}
\end{equation}
where $P_l(x)$ is the Legendre polynomial of order $l$.

To solve Eq.~\eqref{eq:intequation}, a high level of numerical accuracy is required to ensure the systematic cancellation of all divergent terms and the reliable determination of the non-divergent remainder. This is the motivation for the next sections, where we treat our novel method on the two- and three-body level. 

\subsection{Two-body method}\label{Method-2body}
\begin{figure}
    \centering
    \includegraphics{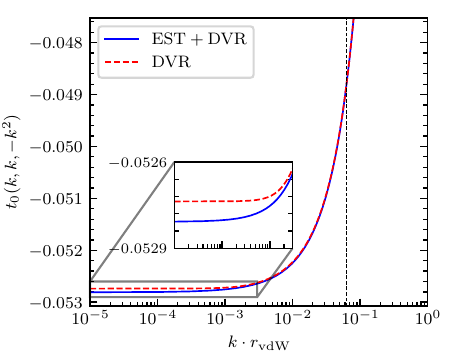}
    \caption{The two-body T-matrix element $t_0(k,k,-k^2)$ for a single-channel van der Waals potential (Eq.~\eqref{eq:LJPotential}) with 1 bound state and a scattering length of $a=-1.05\;r_{\mathrm{vdW}}$, calculated with the DVR-method (red) and our method (blue) described by Eq.~\eqref{eq:Tmatrixswitch}, $p_c$ is indicated with the dashed black line. The inset zooms in on the small momenta and highlights the difference between the DVR and EST method. For this example, the DVR method had a box size of $r_b=1000\;r_{\mathrm{vdW}}$ and for the crossover point we used $C=10$, which resulted in a $\Delta_{\mathrm{cross}}\approx6\cdot10^{-4}$.}
    \label{fig:example_methods}
\end{figure}
The problem of calculating a highly accurate separable expansion was solved in the plane-wave method by using a Weinberg expansion of the interaction potential in free space, thus accurately modeling the low energy scattering behavior of the embedded few-body system \cite{PaperWeinberg}. This approach is unsuitable for our purposes, as the Weinberg expansion is known to converge very slowly for deep molecular interactions \cite{PaulFiniteRange}. Hence, we will instead opt for a direct construction of the two-body transition matrix from eigenfunctions of the interacting two-body Hamiltonian, calculated using a mapped discrete variable representation (DVR) \cite{DVRwilner, ThomasMappedGrid}. Closely related DVR-based techniques have previously been used with great success to compute inelastic three-body observables in alkali-metal systems \cite{ThomasMultiChannel, LiSpinExchange, threebodyJasper}. Unfortunately, since the DVR imposes a hard boundary condition at a large two-body separation, it becomes inaccurate at low energy or momenta, where the two-body wave function becomes sensitive to the box boundary condition. Since we need the zero momentum limit, our novel method solves this limitation of the DVR method by adopting a single-term separable expansion that becomes very precise at this small momentum regime; the Ernst-Shakin-Thaler (EST) expansion \cite{ESTpaper}.

The main advantage of using the EST expansion is its combination of simplicity and accuracy near selected reference energies \cite{ESTpaper, ESTdisadvantage}. At these support points the approximated separable potential is fixed to give the exact same eigenfunctions as the original potential and results in the same solution of the Schrödinger equation. Choosing the support point at $E=0$ means we can use it for low-energy calculations of the two-body transition matrix, since the single-term approximation will remain highly accurate near $k,k'=0$ \cite{PaulFiniteRange}. 

Combining the strengths of both models, we implemented a sharp switch between the two numerical methods at $z_c=-\frac{\hbar^2}{m}p_c^2$. For energies $z<z_c$ we employ the mapped DVR approach, while for higher energies (i.e., closer to zero) in the incoming s-wave channel we use the EST single-term separable approximation, resulting in the following piecewise equation for our two-body transition matrix,  \\
\begin{widetext}
\begin{equation}\label{eq:Tmatrixswitch}
\braket{k'l;\bar{C}'|\mathcal{T}_\alpha(z)|kl;\bar{C}} =t_{l\bar{C}\bar{C}'}(k,k',z)=
    \begin{cases}
    \braket{k';\bar{C}'|\bar{\chi}}\bar{\tau}(z)\braket{\bar{\chi}|k;\bar{C}} & \text{if $(z\geq z_c)\wedge(l=0)\wedge(c=c_{\mathrm{in}})$}\\
    \sum_n\braket{k';\bar{C}'|\chi_{l}^n(z)}\tau_{l}^n(z)\braket{\chi_l^n(z)|k;\bar{C}} & \text{otherwise}
      
    \end{cases}   
\end{equation}
\end{widetext}
where the quantities with an overline indicate the EST form factor and eigenvalue, calculated in the next paragraph. 

As noted in previous work \cite{PaulFiniteRange, ThomasMappedGrid} and confirmed by our own numerical analyses, a good crossover point need only be related to the energy $z$ (and not momenta $k, k'$). Since we expect the DVR method to become inaccurate when the wavelength of the corresponding momentum becomes on the order of the DVR box size $r_b$, a suitable value of the crossover momentum is inversely proportional to $r_b$. So the heuristic choice is made to let the crossover momentum be equal to
\begin{equation}\label{eq:crossovermom}
p_c=C\frac{2\pi}{r_b},
\end{equation}
where $C\geq1$ is a numerical parameter that we tune for optimal accuracy. An example is given in Fig.~\ref{fig:example_methods}, where $t_0(k,k,-k^2)$ is plotted for a single-channel van der Waals potential \cite{PaulFiniteRange}, showing the convergence of both methods close to the crossover point. The transition matrix elements in the figure are off-shell and consider negative energies, because the non-interacting energy of the dimer in the three-body system $Z_c(p)$ is negative. \\

We can calculate the EST form factors and eigenvalues from the condition that the EST potential reproduces the original potential at the support point. This condition means that the potential strength of the incoming (open) channel $\lambda$ is given as $\lambda=\left(\sum_{\bar{C}'}\braket{\psi_{\bar{C}_{\mathrm{in}},E}^{(+)}|V_{\bar{C}_{\mathrm{in}},\bar{C}'}|\psi_{\bar{C}',E}^{(+)}} \right)^{-1}$. Furthermore, the definition of the form factor gives $\braket{k;\bar{C}|\bar{\chi}}=\sum_{\bar{C}'}\braket{k;\bar{C}|V_{\bar{C},\bar{C}'}|\psi_{\bar{C}'}}$. In combination with the support point at $E=0$, the form factor in Eq.~\eqref{eq:Tmatrixswitch} can be calculated as,
\begin{equation}\label{eq:ESTformmultchannel}
\braket{k;\bar{C}|\bar{\chi}}\propto\frac{1}{k}\sum_{\bar{C}'}\int_0^\infty\sin{\left(kr\right)}V_{\bar{C},\bar{C}'}(r)u_{\bar{C}'}(0,r)dr,
\end{equation}
where $u_{\bar{C}'}(0,r)$ is the s-wave two-body radial wavefunction at zero energy in the $\bar{C}'$ channel. We then normalize the form factor such that the open or incoming channel has form factor $\braket{0;\bar{C}_{\mathrm{in}}|\bar{\chi}}=1$. The single eigenvalue $\bar{\tau}(z)$ in Eq.~\eqref{eq:Tmatrixswitch} can be computed using the Lippmann-Schwinger equation and is given by \cite{PaulFiniteRange}
\begin{equation}\label{eq:ESTtau}
\bar{\tau}(z)=\left(\frac{1}{\lambda}-4\pi\sum_{\bar{C}}\int_0^\infty k^2\frac{|\braket{k;\bar{C}|\bar{\chi}}|^2}{z-\frac{\hbar^2k^2}{m}-\varepsilon_{\bar{C}}}dk\right)^{-1},
\end{equation}
where $\varepsilon_{\bar{C}}$ is the two-body channel energy. \\

To assess the quality of the crossover between the two methods, we consider the relative difference between the T-matrices computed using the EST and DVR approaches at the crossover point, defined as 
\begin{equation}\label{eq:accuracycrossover}
\Delta_{\mathrm{cross}}=\frac{t_{\mathrm{EST}}(p_c,p_c,-\frac{\hbar^2}{m}p_c^2)-t_{\mathrm{DVR}}(p_c,p_c,-\frac{\hbar^2}{m}p_c^2)}{t_{\mathrm{EST}}(0,0,0)}.
\end{equation}
From an analytical analysis with square-well potentials in Ref.~\cite{ThomasMappedGrid} and numerically verified for van der Waals-like potentials, it was observed that this relative error scales approximately inversely with scattering length, $\Delta_{\mathrm{cross}}\propto1/|a|$ and with the squared inverse of the box size (and therefore the location of the crossover point in Eq.~\eqref{eq:crossovermom}), $\Delta_{\mathrm{cross}}\propto1/r_b^2$. This means that for a given DVR box size, the scattering hypervolume can be calculated accurately provided that scattering length is larger than some minimal value, $|a|>|a_{\mathrm{min}}|$. In practice, we fix the box size to $r_b=4000$ $r_{\mathrm{vdW}}$ and choose $C=10$. With these parameters, we select the minimum scattering length $|a_{\mathrm{min}}|\sim0.05$ $r_{\mathrm{vdW}}$ for which the crossover error satisfies $\Delta_{\mathrm{cross}}\ll0.01$. \\

\subsection{Implementation and three-body method}\label{Method-3body}
We now consider the numerical implementation of Eq.~\eqref{eq:intequation}, by discretizing the atom-dimer momentum grid and writing it as a matrix equation
\begin{equation}\label{eq:matrixeq}
\mathbf{U}^{\mathrm{nd}}=\mathbf{Z}_0+\underline{Z}\;\underline{\tau}\mathbf{U}^{\mathrm{nd}},
\end{equation}
where $\mathbf{U}^{\mathrm{nd}}$ is the vector form of the non-diverging remainder of $\mathcal{U}^n_{lc}(p_i)$ and where $\mathbf{Z}_0$ is the inhomogeneous term consisting of $\mathcal{I}^n_{lc}(p_i)$ (Eq.~\eqref{eq:inhomterm}). The matrix $\underline{Z}$ contains the kernel function $\mathcal{Z}^{nn'}_{ll';cc'}(p,q)$ (Eq.~\eqref{eq:Kernelfunction}) and $\underline{\tau}$ is the matrix containing the T-matrix eigenvalues $\tau_l^n(Z_c(p_i))\cdot p_i^2\cdot dp_i$. Including multiple partial waves $l$, spin channels $c$, and separable terms $n$ enlarges the vectors and matrices into block form, where each distinct $(l,c,n)$ block uses the same momentum grid, with the exception that the incoming s-wave channel (open channel) employs a momentum grid with double the density to achieve higher accuracy, required for a numerically stable cancellation of the low-momentum divergencies. 

As an order-of-magnitude estimate, a calculation of this system that takes all spin channels and a limited number of partial waves into account, already results in matrices reaching dimensions of order $10^5\times10^5$. Furthermore, these elements are complex-valued, corresponding to more than a hundred gigabytes of memory in double precision. It follows that the main bottleneck of these calculations is the large amount of memory needed \cite{threebodyJasper, threebodyInflation, KraatsThesis}.

To obtain a convergent calculation, we use a maximum partial wave $l_{\mathrm{max}}$, which functions as a cut-off, when the calculation has already converged in terms of higher partial wave contributions \cite{threebodyInflation}. There are also methods for reducing the dimension of the spin basis in order to further reduce the numerical complexity \cite{ThomasMultiChannel, MultichannelEffectsEfimov, spinreductionexample1, spinreductionexample2, spinreductionexample3, spinreductionexample4}. For heavier alkali-metal atoms in particular, it is known that the van der Waals interaction induces a short-range repulsive barrier in the three-body potential that prevents all three atoms being close \cite{universalthreebody1, universalthreebody2}, motivating the Fixed Spectating Spin (FSS) approximation, which we examine in this work \cite{ThomasMultiChannel}. This approximation turns off the interactions when the third particle (spectator) is not in the incoming state, effectively neglecting spin-exchange for this particle, achieved by replacing $V_\alpha$ by $V_\alpha^{\mathrm{FSS}}$, with
\begin{equation}\label{eq:FSS}
V_{\alpha}^{\mathrm{FSS}}=V_\alpha\ket{c_{\mathrm{in}}}\bra{c_{\mathrm{in}}}.
\end{equation}
However, the validity of the FSS approximation relies on the type of atoms, magnetic field range and the probability that they can approach each other in the spin-exchange regime. Refs.~\cite{LiSpinExchange, ThomasMultiChannel} show that for some resonances and zero-crossings the effect of spin-exchange is still significant, which is why we also employ the formal Full Multichannel Spin (FMS) model. Here we use the full possibility of spin-exchange of the third particle, with the number of terms corresponding to the previous order-of-magnitude estimate. This allows us to compare results and see the full effect of spin exchange on the elastic interaction strength of alkali-metal atoms \cite{LiSpinExchange}. \\

\begin{figure}[t]
    \centering  
    \subfloat{}
    \includegraphics{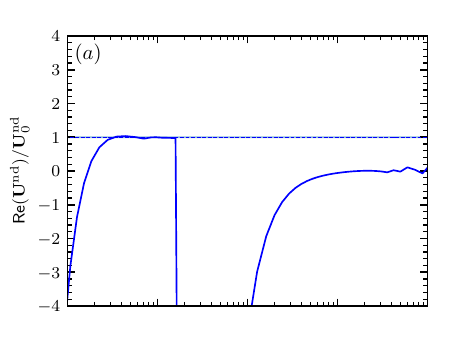}
    \subfloat{}
    \vspace{-8mm}
    \includegraphics{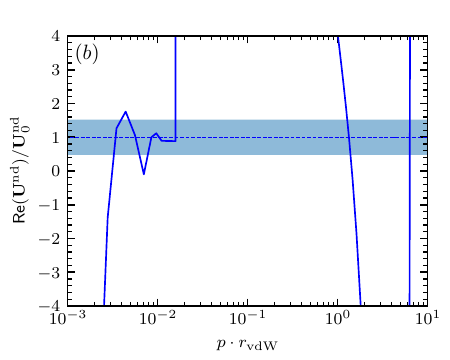}
    \caption{The non-diverging three-body transition operator $\mathbf{U}^{\mathrm{nd}}/\mathbf{U}_0^{\mathrm{nd}}$ as a function of three-body momentum $p$ from a $^{39}\mathrm{K}$ FMS calculation with $M_F=+3$ and $l_{\mathrm{max}}=4$, where $\mathbf{U}_0^{\mathrm{nd}}$ is the value of the plateau around $p\cdot r_{\mathrm{vdW}}\sim0.01$. In (a) the plateau at $a=-0.68\;r_{\mathrm{vdW}}$ is shown and in (b) the plateau at $a=0.74\;r_{\mathrm{vdW}}$ is shown. The dashed line and shaded area corresponds to the average value and standard deviation of points in the plateau respectively.}
    \label{fig:plateau_example}
\end{figure}
\begin{figure*}[t]
    \centering  
    \subfloat{}
    \includegraphics{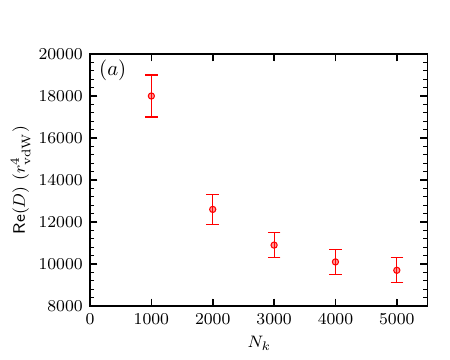}
    \subfloat{}        \includegraphics{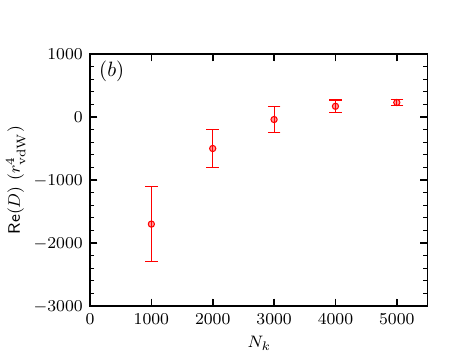}
    \subfloat{}
    \includegraphics{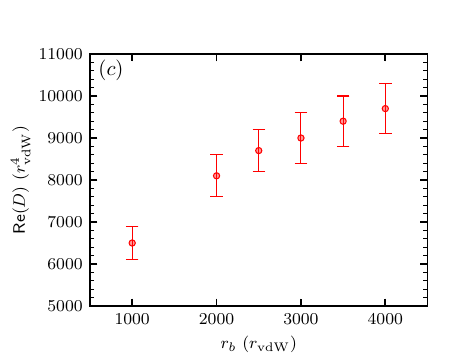}
    \subfloat{}        \includegraphics{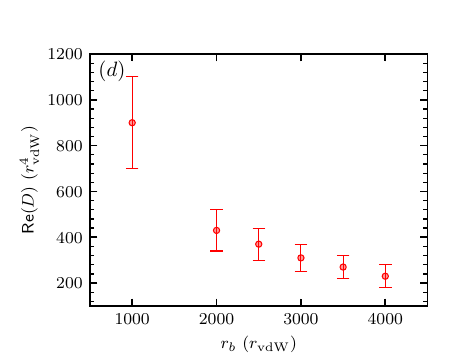}
    \caption{The convergence of the scattering hypervolume at two different scattering lengths from the single-channel LJ-potential in Eq.~\eqref{eq:LJPotential} (with 1 bound state), where (a) and (c) correspond to $a=-1.05\;r_{\mathrm{vdW}}$ and (b) and (d) correspond to $a=0.63\;r_{\mathrm{vdW}}$. Top figures show convergence with respect to the amount of points in the open channel two-body $k$ momentum grid, $N_k$. The bottom figures show convergence with respect to the size of the DVR box $r_b$. The density of DVR points is kept constant meaning that the amount of points scale linear with $r_b$.}
    \label{fig:plateau_convergence}
\end{figure*}
If the cancellation of divergent terms in Eq.~\eqref{eq:intequation} were perfect we should expect the function $\breve{U}_{\alpha0}^{\mathrm{nd}}$ to take on a constant value as the momentum approaches zero, from which we could then calculate the hypervolume using Eq.~\eqref{eq:scathypvol}. In practice however, numerical imperfections in this cancellation leave a remaining divergence in the three-body transition matrix, which will always dominate the hypervolume if the momentum is too small. The challenge then is to increase the accuracy of the calculation to such a degree that there exists a range of low momenta in which $\breve{U}_{\alpha0}^{\mathrm{nd}}$ becomes constant and there exists a plateau where $p$ is small enough, such that polynomial contributions to the transition matrix are damped, while big enough to allow for effective cancellation of the divergences. In Fig.~\ref{fig:plateau_example}, the real value of the non-divergent part of the three-body transition operator $\mathbf{U}^{\mathrm{nd}}$ is shown as a function of momentum $p$ for two different scattering lengths using $^{39}\mathrm{K}$ interaction potentials \cite{TiemannPot}, which for illustrative purposes is normalized such that the value of the plateau $\mathbf{U}^{\mathrm{nd}}_0$ is centered at 1. \\

The quality of a plateau can differ, some are very stable and some can become noisy. This is related to the ratio between scattering length and hypervolume. Recall that a large scattering length also increases the divergent terms in Eq.~\eqref{eq:divergingtransampl}, making it harder to cancel them. This is not directly an issue when the ratio of scattering hypervolume and the strongest diverging term, 
\begin{equation}\label{eq:qualityfactor}
Q=\left.\frac{D}{(2\pi)^6m\hbar^4}/\frac{3a^2}{2\pi^4m\hbar^4p^2}\right|_{p\cdot r_{\mathrm{vdW}}=1},
\end{equation}
remains large. This is the case for Fig.~\ref{fig:plateau_example}a, where $D=3.4(2)\cdot10^3\;r_{\mathrm{vdW}}^4$ and the quality factor is $Q=7.8$. If this quality factor becomes smaller however, there is more noise in the plateau, as is the case for Fig.~\ref{fig:plateau_example}b, which is in the regime where the plateau is least stable. Here $D=450(30)\;r_{\mathrm{vdW}}^4$, but the scattering length remains relatively large and the quality factor reduces to $Q=0.87$. To mitigate the noise, the scattering hypervolume is calculated from the average value of all the points at which their slope is less than an internal tolerance, corresponding to the dashed line in Fig.~\ref{fig:plateau_example}. To measure and include the quality of this plateau and to quantify the uncertainty and noise, we also take the standard deviation of these points, resulting in the shaded areas in Fig.~\ref{fig:plateau_example}. We estimate a 5\% standard uncertainty, attributable to several factors, including convergence of the atom–dimer momentum grid (in density and maximum range), sensitivity to other numerical parameters, and the choice of cutoff in the separable approximation of the two-body T-matrix (Eq.~\eqref{eq:tmatrixexpansion}). If the standard deviation of a plateau exceeds the standard uncertainty of 5\%, we take this as an improved estimate of the uncertainty. This uncertainty is used to define the error bars in Fig.~\ref{fig:plateau_convergence} and in all figures where the scattering hypervolume is shown. Systemic uncertainties, such as accuracy of the interaction potentials, are excluded. Additionally, for a more detailed discussion of the effects of separable approximations on studying three-body physics, we refer to Ref.~\cite{PaulFiniteRange}.  

When $p$ is further increased, another noticeable feature is a sudden discontinuity at $p_c$ (here $p_c\cdot r_{\mathrm{vdW}}=1.57\cdot10^{-2}$). This is where we switch to the DVR-method as illustrated in Fig.~\ref{fig:example_methods} and Eq.~\eqref{eq:Tmatrixswitch}. It is important to note that $\breve{U}_{\alpha0}^{\mathrm{nd}}$ should not be confused with the actual full three-body transition matrix. A discontinuity is expected because only the divergences in the first separable term are subtracted (sufficient for the EST single-term expansion), so that in the DVR regime the subtraction removes contributions that do not correspond to the actual divergences. 

Two parameters improve the canceling of diverging terms for smaller three-body momenta; the two-body (dimer) momentum grid $k$, on which we compute the two-body transition matrix, and the size of the DVR box $r_b$. The size of the DVR box is important, because it calculates the zero-energy wavefunction needed as input in the EST method in Eq.~\eqref{eq:ESTformmultchannel}. A larger grid in turn generates a wavefunction that is closer to the free space wavefunction and therefore makes the calculation of the EST form factor and eigenvalue more accurate. For both parameters the computation time in calculating and diagonalizing the two-body $T$-matrix scales cubically. We have calculated the convergence of the scattering hypervolume with respect to the aforementioned parameters and the results are presented in Fig.~\ref{fig:plateau_convergence}. This data motivated us to use a dimer $k$-grid of 5000 points in the incoming (open) channel and 1000 in the higher partial waves and spin-exchange (closed) channels (smaller contribution and less resources needed), and a DVR grid with size $r_b=4000\;r_{\mathrm{vdW}}$. This is a doubling of the grid size as compared to those used in Refs.~\cite{ThomasMappedGrid, ThomasMultiChannel} and results in the plateaus in Fig.~\ref{fig:plateau_example}. 

\subsection{Benchmarking}\label{Method-benchmarking}
\begin{figure}
    \centering
    \includegraphics{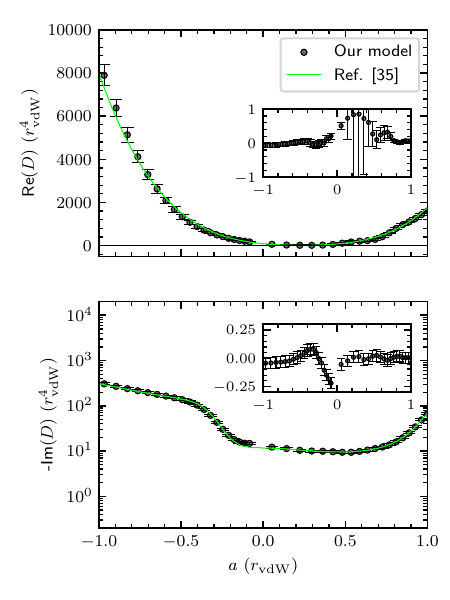}
    \caption{The real and imaginary part of the scattering hypervolume $D$ from a single channel LJ-potential (with 1 bound state) calculation, with maximum partial wave $l_{\mathrm{max}}=16$. This is compared to the results from the method in Ref.~\cite{vdWuniversality}, representing the green lines. The insets show the relative difference between both methods as a function of $a$, expressed as a dimensionless ratio.}
    \label{fig:singlechannelD}
\end{figure}

To validate our approach against the plane-wave method of Ref.~\cite{scathypPaul, vdWuniversality} in the single-channel, shallow potential limit, we perform benchmark calculations for a Lennard-Jones (LJ) interaction. The same potential is used for the results shown in Fig.~\ref{fig:plateau_convergence} and is defined as
\begin{equation}\label{eq:LJPotential}
V_{\mathrm{LJ}}(r)=-\frac{C_6}{r^6}\left(1-\frac{\sigma^6}{r^6}\right),
\end{equation}
where $\sigma$ is the distance at which the potential turns repulsive and sets the number of allowed bound states and the value of the two-body scattering length $a$. Our calculations of the three-body scattering hypervolume $D$ as a function of the two-body scattering length $a$ for a LJ-potential, with one s-wave bound state ($\sigma$ between resonances at $\sigma=0.920\;r_{\mathrm{vdW}}$ and $\sigma=0.573\;r_{\mathrm{vdW}}$) and $l_{\mathrm{max}}=16$ are presented in Fig.~\ref{fig:singlechannelD}. 

The insets in Fig.~\ref{fig:singlechannelD} show the relative difference between the two methods. They demonstrate that, within the error margins, most of our results are consistent with those of Ref.~\cite{vdWuniversality}. The cause of the relatively large mismatch in $\mathrm{Re}(D)$ at small and positive $a$ is discussed in the previous Sec.~\ref{Method-3body}. Here it is demonstrated that it is more difficult to compute the hypervolume in this regime with our method, as is reflected in the larger error bars. When $\mathrm{Re}(D)$ approaches zero, the evaluation of the relative error might also suffer from round off errors. The relative mismatch of $\mathrm{Im}(D)$ between the methods is averaged over all the data points, giving an average relative mismatch of 4\% and showing that the estimated error of 5\% is reasonable. However, the area at small and negative $a$ shows a larger mismatch. This can be attributed to either the discussion in Sec.~\ref{Method-2body} about our two-body method becoming less accurate at vanishing $a$, or due to uncertainties in the method of Ref.~\cite{vdWuniversality}. This work only partly shows the uncertainty of their results, but it is of similar scale as our own.

In Ref.~\cite{vdWuniversality} it was pointed out that the real part of the scattering hypervolume behaves universally for weakly interacting systems with van der Waals potentials. This behavior was attributed to the characteristic strong repulsive barrier in the three-body potential, which may be modeled by a hard-hypersphere interaction \cite{vdWuniversality, IncaoFewBody, BraatenHammerBEC}. In the weakly interacting regime the scattering length cannot be the only length scale to determine the location of the barrier. Here, the hard hypersphere radius $R_{\mathrm{hh}}$ is given as $R_{\mathrm{hh}}=|a-a_{\mathrm{hh}}^{\pm}|$, with $a_{\mathrm{hh}}^\pm$ an offset that captures finite range effects, and where the + (-) sign is the offset for the hypervolume at positive (negative) scattering lengths. Hard-hypersphere scattering predicts a universal scaling of $D$ as
\begin{equation}\label{eq:hard-hypersphere}
\text{Re}(D)=C_{\pm}R_{\mathrm{hh}}^4=1689|a-a_{\mathrm{hh}}^{\pm}|^4,
\end{equation}
where $C_{\pm}$ is the scaling factor \cite{vdWuniversality}. The universal van der Waals results of Ref.~\cite{vdWuniversality} result in finite range offsets of $a_{\mathrm{hh}}^+=-0.010(3)\;r_{\mathrm{vdW}}$ and $a_{\mathrm{hh}}^-=0.474(7)\;r_{\mathrm{vdW}}$. From fitting Eq.~\eqref{eq:hard-hypersphere} to the data from our model we extract the location of the hard-hypersphere barrier and obtain the following finite range offsets; $a_{\mathrm{hh}}^+=0.001(5)\;r_{\mathrm{vdW}}$ and $a_{\mathrm{hh}}^-=0.466(5)\;r_{\mathrm{vdW}}$, showing good agreement with the results previously obtained through the plane-wave method.

Due to the universality, it is expected that these results are generally valid for alkali-metal gases, provided that the Feshbach resonance is relatively broad and the effects of three-body spin-exchange are weak, therefore any deviation from these values must likely arise from multichannel effects.

\section{Results}\label{Results}
\begin{figure}
    \centering
    \includegraphics{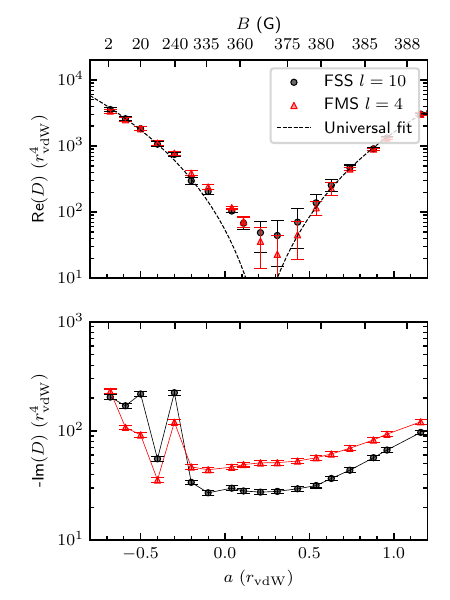}
    \caption{The scattering hypervolume for $^{39}\mathrm{K}$ in the state $\ket{f=1,m_f=1}$, from two different calculations, FSS with $l_{\mathrm{max}}=10$ and FMS with $l_{\mathrm{max}}=4$ together with the universal behavior, which is a fit of Eq.~\eqref{eq:hard-hypersphere}, but with different scaling constants. This graph shows the real and imaginary part of the three-body scattering hypervolume versus the two-body scattering length, with the corresponding magnetic field strengths on the top axis. }
    \label{fig:PotassiumD+}
\end{figure}
\begin{figure}
    \centering
    \includegraphics{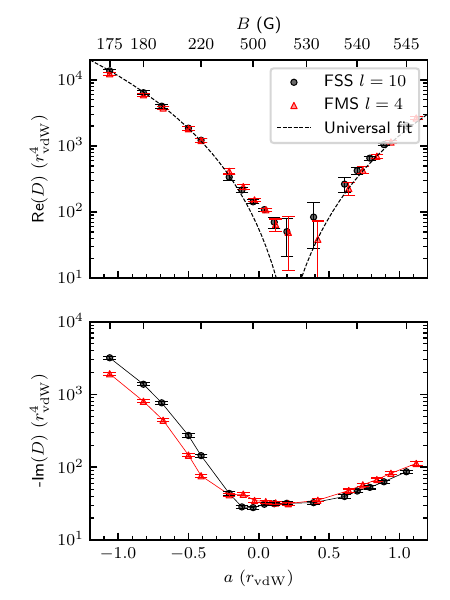}
    \caption{The scattering hypervolume for $^{39}\mathrm{K}$ in the state $\ket{f=1,m_f=-1}$, from two different calculations, FSS with $l_{\mathrm{max}}=10$ and FMS with $l_{\mathrm{max}}=4$ together with the universal behavior, which is a fit of Eq.~\eqref{eq:hard-hypersphere}, but with different scaling constants. This graph shows the real and imaginary part of the three-body scattering hypervolume versus the two-body scattering length, with the corresponding magnetic field strengths on the top axis. }
    \label{fig:PotassiumD-}
\end{figure}
We now turn to the central application of our method: extracting the three-body scattering hypervolume for a realistic alkali-metal system with full internal state structure. We focus on ultracold, spin-polarized $^{39}\mathrm{K}$, whose three-body properties have been extensively explored both theoretically \cite{ThomasMultiChannel} and experimentally \cite{PotassiumexperimentEigen, cambridgepaper, Arlt39-K1, Arlt39-K2}. Specifically, we consider spin-polarized ensembles in the $f=1$ manifold, with all atoms prepared in the $\ket{f=1,m_f=1}$ or $\ket{1,-1}$ hyperfine state. In each case the three-body problem is restricted to a fixed conserved total projection, $M_F=+3$ or $M_F=-3$, which defines the coupled-channels basis used in our calculations. The associated two-body scattering channel in the $M_F=3$ system is known to posses a zero-crossing of the s-wave two-body scattering length at $B\approx 350$ G, and the $M_F=-3$ system has a zero-crossing at $B\approx506$ G, on which we focus for our study of the hypervolume \cite{PotassiumResonance1, PotassiumResonance2}. We have performed FMS calculations with a maximum partial wave $l_{\mathrm{max}}=4$ and FSS calculations with $l_{\mathrm{max}}=10$, in similar vein to previous work in Refs.~\cite{ThomasMultiChannel, LiSpinExchange}. 

The scattering hypervolume for both calculations in the weakly-interacting regime is presented in Figs.~\ref{fig:PotassiumD+} and \ref{fig:PotassiumD-}, together with the associated magnetic field ranges. The overlap between the FMS and FSS simulations for the real part of the scattering hypervolume in both systems shows that when $^{39}\mathrm{K}$ atoms elastically scatter they do not reach the short range where spin-exchange becomes prominent This hints to an effective repulsive three-body potential barrier, in accordance with the predicted universal behavior for long-range van der Waals interactions \cite{universalthreebody1, universalthreebody2}. However, our results suggest that the imaginary part of $D$ in both systems is more sensitive to the truncation of the three-body spin basis than the real part. This is generally to be expected, since elastic scattering is long range and inelastic scattering involves dimer states which are bound at short distances and are therefore more sensitive to the spin-exchange effects. \\

Assuming that the hard-hypersphere scattering model provides an adequate description also in the multichannel case, we again fit Eq.~\eqref{eq:hard-hypersphere} to $\text{Re}(D)$ in the low-$a$ regime. We find that $^{39}\mathrm{K}$ follows the expected universal $a^4$ scaling in the weakly interacting limit, as indicated by the dashed black lines in Figs.~\ref{fig:PotassiumD+} and \ref{fig:PotassiumD-}, but with different prefactors for $a>0$ and $a<0$. For positive $a$, this universal scaling follows $\text{Re}(D)_{+3}\approx1.9(2)\cdot10^3|a-0.04(3)|^4$ and $\text{Re}(D)_{-3}\approx1.56(6)\cdot10^3|a+0.02(1)|^4$, where $\pm3$ indicates which $M_F$ system is concerned. For negative $a$, the universal scaling follows $\text{Re}_{+3}(D)\approx3.2(3)\cdot10^3|a-0.36(2)|^4$ and $\text{Re}_{-3}(D)\approx3.4(1)\cdot10^3|a-0.358(7)|^4$. These results demonstrate that a universal scaling is present and quite similar for both spin systems, but that the finite-range parameters have changed. Most notably, the location of the hard-hypersphere barrier for negative $a$ shifts from $a_{\mathrm{hh-LJ}}^-=0.466(5)\;r_{\mathrm{vdW}}$ to $a_{\mathrm{hh-39K}}^-=0.36(2)\;r_{\mathrm{vdW}}$ for both spin systems and the scaling factor $C_{-}$ changes from $1689$ to $3200$ and $3400$. We hypothesize that this behavior arises from the multichannel character of our calculations. It is noted that the three-body parameters, depth and shape of the effective three-body potential are dependent on the type and width of the multichannel resonance \cite{Multchannelhyperspherical, EfimovianJasper, Reshapedthreebodypot}. For the $^{39}\mathrm{K}$ zero-crossing under current investigation, the resonance width of the nearest resonance is $s_{\mathrm{res}}=2.1$, which is in the broad to intermediate range \cite{PotassiumResonance1}. In this regime, the two-body effective range and three-body parameter behave non-universally and the depth of the hyperspherical potential is shallower than for a single-channel system \cite{effectiverangeexp, EfimovianJasper, Reshapedthreebodypot}. These deviations in the effective three-body interaction potential from the universal single-channel case could provide the mechanism for the shift in the location of the effective hard-hypersphere barrier and the stronger scaling of the scattering hypervolume. \\

\begin{figure}[t]
    \centering
    \includegraphics{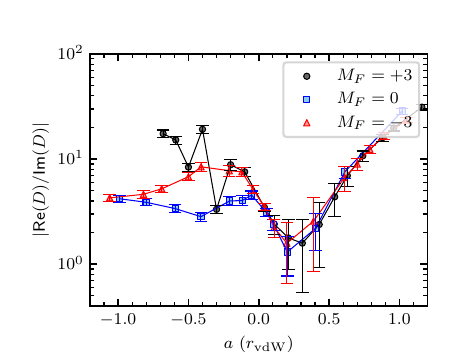}
    \caption{The ratio of the real and imaginary part of $D$, plotted for three initial spin states with different total spin quantum number $M_F$. These results are obtained using the FSS model with $l_{\mathrm{max}}=10$.}
    \label{fig:hypervolumefrac}
\end{figure}

In Fig.~\ref{fig:hypervolumefrac}, the ratio between the real and imaginary part of the scattering hypervolume of $^{39}\mathrm{K}$ is shown, now also with the initial $\ket{1,0}$ state (calculated using the FSS model near the resonance at $B=471$ G). Fig.~\ref{fig:hypervolumefrac} shows how the ratio is generally consistent for all considered initial spin states, apart from some noticeable differences. \\
These differences are highly interesting, because to observe elastic three-body scattering experimentally, a large fraction between elastic and inelastic scattering at small scattering length is needed, such that the BEC has a long enough lifetime \cite{cambridgepaper}. From Fig.~\ref{fig:hypervolumefrac}, we note that $^{39}\mathrm{K}$ spin polarized in the $\ket{1,1}$ state is the most promising candidate for experimental research on three-body interactions, with a large ratio of elastic to inelastic scattering. \\

\begin{figure}[t]
    \centering
    \includegraphics{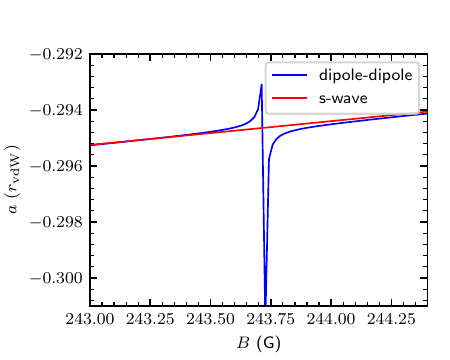}
    \caption{The two-body scattering length for a purely s-wave calculation (red line) and a calculation including dipole-dipole interactions, showing the appearance of a d-wave resonance.}
    \label{fig:twobodyresonance}
\end{figure}
A notable deviation occurs for the $M_F=+3$ system: in Fig.~\ref{fig:PotassiumD+} the data point at $a\approx-0.3\;r_{\mathrm{vdW}}$ (corresponding to $B=240$ G) lies slightly above the universal curve. A related feature appears in the inelastic contribution, where the deviation is more pronounced and leads to a minimum in the ratio $\text{Re}(D)/\text{Im}(D)$ shown in Fig.~\ref{fig:hypervolumefrac}. Further investigation reveals that at $B = 243.7$ G a d-wave dimer state appears in the incoming two-body channel, with a two-body spin projection $m_{\mathrm{2b}} =+2$. This d-wave resonance, present in both FSS and FMS models, enacts a sharp increase of the three-body recombination rate. The resonance is also seen in a two-body scattering calculation that includes anisotropic dipole-dipole interactions, visible in Fig.~\ref{fig:twobodyresonance}. Since the anisotropic dipole-dipole interaction gives a typically subdominant contribution to the two-body scattering physics, the resonance at the two-body level is very narrow. This is very different to the three-body level, where we observe a much wider resonance, with its signature appearing already at an s-wave scattering length $a=-0.3\;r_{\mathrm{vdW}}$ or magnetic field of $B=238$ G in Fig.~\ref{fig:PotassiumD+}. This is likely related to the d-wave resonance at the three-body level having a different character and appearing within the typically stronger isotropic van der Waals interaction. On the three-body level, the resonance is now accessible due to the three-body geometric coupling of partial waves, as seen in the kernel function in Eq.~\eqref{eq:Kernelfunction}. It is interesting that we are now able to see the small effect of such a resonance on the elastic scattering strength. Furthermore, the fact that we can recognize this bound state in our three-body simulation showcases the correct embedding of the molecular two-body physics in our method.

We also note recent work on three-body systems with both van der Waals and dipole-dipole interactions, which examined the impact of dipolar interactions on universal three-body properties \cite{OiEndo}. Although such anisotropic interactions lie beyond the scope of the present work, they represent an interesting direction for future extensions of our framework.

\section{Conclusion}\label{Concl}
In conclusion, we have developed and demonstrated a novel coupled-channels method for elastic three-body scattering that enables the extraction of the three-body scattering hypervolume $D$ for realistic alkali-metal interactions, including deep multichannel molecular potentials with many bound states. We applied the method to spin-polarized $^{39}\mathrm{K}$ and obtained the complex hypervolume with controlled numerical accuracy. 

Our multichannel results recover the expected universal $a^{4}$ scaling of $\text{Re}(D)$ in the weakly interacting regime, as previously established for single-channel van der Waals model potentials. At the same time, we find systematic quantitative differences: the prefactor on the negative-$a$ side is comparatively larger, and the effective finite-range parameter $a_{hh}^{-}$ is reduced relative to the single-channel reference. These deviations are consistent with the notion that the intermediate width of the relevant nearby Feshbach resonance at the employed zero crossings can modify the effective three-body potential and thereby induce departures from simple universality. In addition, by analyzing the influence of a two-body $d$-wave dimer, we show that the present framework incorporates the relevant two-body physics at the required level of detail and makes explicit how two-body resonances can imprint themselves on three-body observables. 

This work opens several directions for further research. From an experimental perspective, quantitatively reliable multichannel predictions are particularly valuable because clear signatures of elastic three-body interactions have not yet been observed directly. The predicted stabilization of homogeneous quantum droplets by elastic three-body interactions \cite{bedaquedrops, bulgacdrops, zwergerdrops, PetrovControl} provides a promising route to access genuine three-body contributions to the phase diagram of dilute quantum gases, provided one can reach a regime of small scattering length and sufficiently large $\text{Re}(D)/\text{Im}(D)$ \cite{vdWuniversality, cambridgepaper}. In light of our results, the $\ket{f=1,m_f=+1}$ state of $^{39}\mathrm{K}$ appears especially promising in this regard. 

More broadly, by enabling quantitative coupled-channels predictions across different spin configurations (and for other atomic species, such as lithium \cite{lithiumexperiments2, lithiumexperiments1}) our approach supports targeted searches for zero crossings and magnetic-field regions that maximize the ratio of elastic to inelastic three-body scattering. In this way, the method can provide concrete guidance for experiments by identifying parameter regimes where elastic three-body effects should be large enough to become observable, for example through their influence on BEC stability and collapse dynamics \cite{cambridgepaper, Collapsedynamics}.\\

\begin{acknowledgments}
We thank Christoph Eigen, Jing-Lun Li, Raul dos Santos and Jasper Postema for fruitful discussions. J.v.d.K.and S.K. acknowledge financial support from the Netherlands Organization for Scientific research (NWO) under Grant No. 680.92.18.05, and from the Dutch Ministry of Economic Affairs and Climate Policy (EZK), as part of the QDNL program. D.A.-B. acknowledges funding from the Research Foundation-Flanders via a postdoctoral fellow ship (Grant No. 1222425N). The results presented in this work were obtained on the TU/e Umbrella HPC cluster of the Eindhoven Supercomputing Center. 
\end{acknowledgments}

\bibliography{References}

\end{document}